\documentclass[nonacm, sigconf]{acmart}

\renewcommand\footnotetextcopyrightpermission[1]{} 
\usepackage{float}
\usepackage{booktabs} 
\usepackage{times}
\usepackage{graphicx}
\usepackage{amsmath}
\usepackage{enumitem}
\DeclareMathSizes{12}{30}{16}{12}

\makeatletter
\renewcommand\@formatdoi[1]{\ignorespaces}
\makeatother
\setcopyright{rightsretained}
\acmDOI{ }
\acmISBN{}
\acmConference[RecSysKTL'18]{Workshop on Intelligent Recommender Systems by Knowledge Transfer and Learning}{October 2018} {Vancouver, Canada}
\acmYear{2018}
\copyrightyear{2018}
\acmArticle{}
\acmPrice{}

\begin{document}
\title{A Hybrid Variational Autoencoder for Collaborative Filtering}

\author{Kilol Gupta}
\affiliation{%
  \institution{Columbia University}
  \city{New York}
  \state{NY}
  \postcode{10027}
}
\email{kilol.gupta@columbia.edu}

\author{Mukund Y. Raghuprasad}
\affiliation{%
  \institution{Columbia University}
  \city{New York}
  \state{NY}
  \postcode{10027}
}
\email{my2541@columbia.edu}

\author{Pankhuri Kumar}
\affiliation{%
  \institution{Columbia University}
  \city{New York}
  \state{NY}
  \postcode{10027}
}
\email{pk2569@columbia.edu}


\begin{abstract}
In today's day and age when almost every industry has an online presence with users interacting in online marketplaces, personalized recommendations have become quite important. Traditionally, the problem of collaborative filtering has been tackled using Matrix Factorization which is linear in nature. We extend the work of \cite{liang2018variational} on using variational autoencoders (VAE) for collaborative filtering with implicit feedback by proposing a hybrid, multi-modal approach. Our approach combines movie embeddings (learned from a sibling VAE network) with user ratings from the Movielens 20M dataset and applies it to the task of movie recommendation. We empirically show how the VAE network is empowered by incorporating movie embeddings. We also visualize movie and user embeddings by clustering their latent representations obtained from a VAE.
\end{abstract}

%
%



\maketitle

\section{Introduction \& Related Work}
Recommender systems have become important in today's landscape where social media and online interactions have grown. People frequently make choices with regard to the news articles they read, the products they buy, songs they listen to and the movies they watch with the help of recommender systems. All of these applications have potential new products to be discovered by users. When combined with personalized recommendations, it leads to increased user engagement, satisfaction, and increased business profits. The task of generating personalized recommendations has historically been and continues to be challenging. Essentially, the task is to recommend items to users, based on user context such as view history, click-through rate, demographic information etc. and context information on items such as popularity, genre, description, reviews etc.\\
Collaborative Filtering (CF) has been one of the most widely used methods. The model-based CF includes techniques like Latent Factor Models such as Matrix Factorization \cite{koren2009matrix, mnih2008probabilistic}. However, these methods are linear in nature whereas the interaction between users and items is seemingly non-linear.\\
Neural Networks have made remarkable progress in achieving encouraging results in digital image processing \cite{krizhevsky2012imagenet}, natural language processing \cite{collobert2011natural}, speech recognition \cite{hinton2012deep} and autonomous driving \cite{al2017deep}. Neural networks (deep learning) has proven successful because of its ability to model complicated non-linear data representations \cite{bengio2013representation}. The aforementioned CF algorithms try to generate a latent representation of the interaction between user and items. Better characterization of this interaction will supposedly lead to better recommender systems. There has been promising work of applying deep learning to the field of collaborative filtering \cite{lee2017augmented, zheng2016neural, wu2016collaborative, he2017neural}.\\
Variational Autoencoders (VAEs) have recently been adapted for the task of personalized recommendation \cite{liang2018variational}. Our paper draws motivation from this work to empirically study if augmenting movie ratings with movie embeddings result in a better characterization of the interaction between users and items (movies in this case). We do so by first using a VAE network to learn movie embeddings and then augmenting the user ratings with these. This mixed representation is then fed into a second VAE network that learns from a collaborative filtering model. We call this new network as Hybrid VAE (Fig. 2). For comparison, we start by implementing a standard VAE (Fig. 1). Overall, this paper attempts to assess the implementation, applicability, merits, and overhead of a hybrid VAE for the task of collaborative filtering.

\section{Dataset}
\textbf{MovieLens 20M dataset \cite{harper2016movielens}}: This dataset contains 20,000,263 ratings across 27,278 movies as generated by 138,493 users. We randomly divide the set of users into training, test and validation sets with 10,000 users each in test and validation and 118,493 in training. To standardize our input, we discard those movies that did not have any information on IMDb\footnote{https://www.imdb.com/ \label{footnote 1}}, leaving a total of 26,621 movies.\\
The ratings are binarized with 1 for movies that the user had rated greater than 3.5, and 0 otherwise. The threshold of 3.5 is chosen to be consistent with \cite{liang2018variational}. Binarization offers an elegant way to fairly treat the unseen movies as belonging to class 0 (implicit feedback). The VAE outputs a probability distribution over the list of movies for each user and the loss function optimizes the difference between the outputted probability and the binarized user rating. The expectation is that the trained model will output a probability as close to 0 as possible for a movie with the binary rating of 0 and a probability as close to 1 for a movie with the binary rating of 1. Such a direct correlation between the network input (ratings) and the output (probability) isn't possible if original ratings on a scale of 0 to 5, are used.

\section{Evaluation Methodology}
We create 3-fold cross-validations (CVs) of the dataset to ensure robustness of results. All the results reported in this paper are averaged over the 3 CVs. We find the standard deviation across CVs to be in the order of $10^-3$. We use the rank-based evaluation metrics: Recall@20, Recall@50, and NDCG@100. Similar to \cite{liang2018variational}, for each user in the test set, we compared the predicted ranks of the movies with their true ranks. The predicted ranks were decided by sorting the output of the last layer of the VAE network gives a probability distribution on the movies. While Recall@R considers all items ranked within the first R items to be equally important, NDCG@R uses a monotonically increasing discount to emphasize the importance of higher ranks versus lower ones. Formally, $w(r)$ is defined as the item at rank $r$, $I[\ ]$ is the indicator function, and $I_u$ is the set of held-out items that user u clicked on. Then Recall@R for user u is defined as:
\begin{equation}
    Recall@R(u, w) = \frac{\sum_{r=1}^R I[w(r) \in I_u]}{min(M, |I_u|)}
\end{equation}
We choose the minimum of R and number of items clicked by user u as our denominator to normalize Recall@R, which corresponds to ranking all relevant items in the top R positions. The truncated, discounted cumulative gain (DCG@R) is defined below. NDCG@R is the DCG@R linearly normalized to $[0, 1]$, after dividing it with the best possible DCG@R, where all the held-out items are ranked at the top.
\begin{equation}
    DCG@R(u.w) = \sum_{r=1}^R\frac{2^{I[w(r) \in I_u]} - 1}{log(r+1)}
\end{equation}
Two types of evaluation schemes are used: 
\begin{itemize}
\itemsep0em
\item Eval 1: The first scheme is where the training is performed on 118,400 users with testing on 10,000 users and validation on 10,000 users. The evaluation metrics NDCG and Recall are then computed for each test user over all of the 26,621 movies.  
\item Eval 2: The difference from Eval 1 is that the click history of each test user is divided into an 80/20 split. The binarized rating of the movies in the $20\%$ split is set to 0 and the remaining $80\%$ split is left unchanged. NDCG@k and Recall@k are then calculated for each test user considering only the $20\%$ split. This scheme is stricter and closer to the real world as it evaluates the prediction of the model on movies which are previously unseen by the user.
\end{itemize}

\section{Movie Feature Extraction}
The information from an auxiliary source to the primary user-rating information is fed to the original VAE network as an item-embedding.
The feature extraction is done using three information sets: movie genres, genome tags and features from the IMDb movie summaries.
\subsection{Movie genres}
The movie genres provided in the MovieLens-20M dataset are used for this category. The dataset categorizes the movies into a set of 19 genres and each movie can belong to multiple genres. A binary encoding of all genres for each movie is created as a feature vector for a movie.
\subsection{Genome tags}
The MovieLens-20M contains predetermined genome tags ranging from characteristics of the plot to actors in the movies. A few examples of these genome tags are "based on a book", "computer animation", "1920" and etc. A total of 1128 tags have been used in this dataset, where, each movie is tagged with multiple tags and a relevance score is associated with each movie-tag pair. For this paper, the top 20 tags for each movie are considered and a binary vector of these top tags is created as the feature vector.
\subsection{IMDb summary}
The data for this category is collected using the Online Movie Database API \footnote{http://www.omdbapi.com/ \label{footnote 2}} and features are extracted for 26,621 movies. Each movie is associated with a language, a movie certification, a viewer review score/IMDb rating and, the movie plot. For feature extraction, we use the following information:
\begin{itemize}
\itemsep0em
\item Language: A one-hot encoding for the language of the movie is created using all the languages mentioned in the dataset.
\item Certification: This is a one-hot encoding for the certification given to the movie. Example: PG-13, R etc.
\item IMDb rating: The score given is a continuous value that ranges from 0 to 10. This is incorporated directly without any transformation. 
\item Plot: The plot is analyzed and various features, describing different aspects of the text are extracted. 
\begin{itemize}
\itemsep0em
\item Linguistic Inquiry and Word Count (LIWC) \cite{liwc}: LIWC is a lexicon dictionary which associates words in the English language to linguistic, psychological and sociological processes into 64 categories. The plot text is tokenized and each token is mapped to a binary vector of the LIWC category it belongs to. The average vector for the whole plot is obtained by averaging the 64-d binary vectors for all the tokens in the plot. 
\item Valence Arousal and Dominance(VAD) \cite{warriner2013norms}: VAD is also a lexicon dictionary that associates words with a 3-d score. The plot text is tokenized and scores are averaged for all words in the movie plot. 
\item Word2Vec \cite{mikolov2013efficient}: To capture the semantic differences and similarities between movies and their plots, the averaged Word2Vec vector for the plot text is also used as a feature. Pre-trained 300-d Word2vec is used.  
\end{itemize} 
All the above-mentioned features are then concatenated to give a 671-d movie feature vector.
\end{itemize}

\section{Implementation Details}
\subsection{Standard-VAE}
\begin{center}
\includegraphics[scale=0.6]{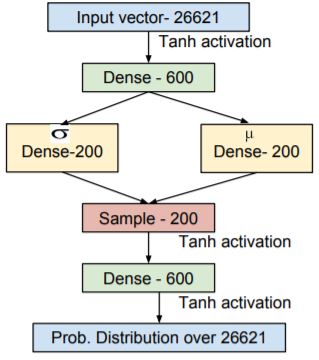}
\captionof{figure}{Standard VAE architecture}
\end{center}

The Standard-VAE considered in this paper takes user ratings $x_u$ as input. The user input is encoded to learn the mean, $m_u$ and the standard deviations $\sigma_u$ of the $K$-dimensional latent representation through the encoder function $g_\phi()$ (3) . The latent vector for each user, $z_u$ is sampled using $m_u, \sigma_u$. The decoder function $f_\theta()$ (4) is then used to decode the latent vector from $K$-dimensions to a probability distribution $\pi_u$ in the original $N$-dimension. This distribution gives us the probability of the $N$-movies being viewed by user $u$. 
\begin{equation}
    g_\phi(x_u) = m_u, \sigma_u\ \ \  z_u \sim N(m_u, \sigma_u)
\end{equation}
\begin{equation}
    f_\theta(z_u) = \pi_u
\end{equation}
The standard-VAE in this paper differs from the normal VAE which has the final output as the reconstructed input. Here, the output is a probability distribution over the K items. The objective function/loss used in the model is the ELBO\cite{liang2018variational} given in (5).
\begin{equation}
loss = \log p_\theta(x_m|z_m) + KL(q(z_m)||p(z_m|x_m))
\end{equation}
Where, $x_m$ is the movie feature vector while $z_m$ is its latent representation. Here, the first part of the equation considers the log-likelihood for a movie given its latent representation and the second part is the Kullback-Leibler (KL) divergence measure. The log-likelihood function considered is given as,
\begin{equation}
\begin{aligned}
\log p_\theta(x_u|z_u)=\sum_i x_{ui} \log \sigma(f_{ui}) + (1-x_{ui})\log (1-\sigma(f_{ui}))
\end{aligned}
\end{equation}
where $\sigma(x) = 1/(1 + \exp(-x))$ taken over all the items $i$. 
The KL Divergence is calculated for the latent state of the model, $z_u$.

\subsection{Hybrid-VAE}
\begin{figure}[H]
\includegraphics[scale=0.5]{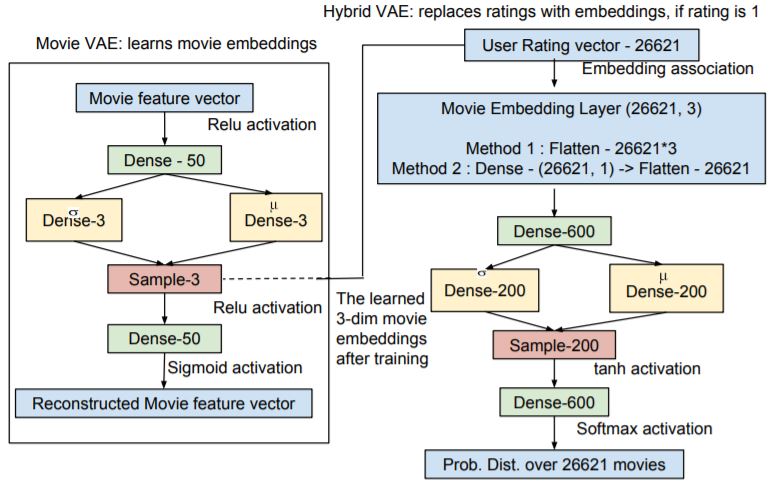}
\captionof{figure}{Hybrid VAE architecture}
\end{figure}
Incorporating a high-dimensional feature vector for each movie into an already high dimensional user-rating input is computationally difficult. For this, a Movie-VAE (M-VAE) is used to encode the movie feature vectors into a low-dimensional latent space.
The M-VAE is trained over the movie features extracted from all 26,621 movies. The size of the latent space considered is equal to the dimension of the movie embeddings. This paper considers a movie embedding of size 3. The movie features are encoded into these embeddings and used in the Hybrid-VAE network.\\

The Hybrid-VAE (H-VAE) is similar to the Standard VAE, but it contains an extra layer which combines movie embeddings with the user ratings for each movie. The embeddings are obtained from the Movie-VAE (M-VAE). Indices of movies are maintained across the M-VAE and H-VAE so that embeddings and ratings can be matched. Movies with user-click history as 0 are assigned a zero-embedding (a 3-dim 0 vector). The architecture for the Hybrid-VAE is given in Fig. 2. Given a user click history $x_u$, the embedding input $x'_u$, for each movie $i$, is given by, 
\begin{equation}
\begin{aligned}
&\ \ \ \ \ \ \ x'_u = <e_1, e_2, e_3,..., e_n>,\\
e_i =& 
\begin{cases} 
    Movie\_Embedding(i),\ if\ x_{ui} = 1,\\
    Movie\_Embedding(0),\ if\ x_{ui} = 0  
\end{cases}
\end{aligned}
\end{equation}
The successive steps follow the same procedure as the Standard-VAE, with $x'_u$ replacing $x_u$ in the encoding procedure. However, the objective function still considers the input user-click history $x_u$ instead of the embedding input $x'_u$. It is worth noting that the embedding layer output in the H-VAE is a 3D matrix of dimensions, \\$(batch\ size \times num\ of\ movies \times movie\ embedding\ dimension)$. The intermediate dense layer, however, requires a 2D vector as input. There are two ways to introduce the embeddings into the intermediate layer:
\begin{itemize}[noitemsep,topsep=0pt]
\item Flatten the 3D embedding layer output into a 2D layer: In this case, the input to the intermediate dense layer is a vector of length equal to\\ $number\ of\ movies \times movie\ embedding\ dimension$
\item Convert the 3D embedding into a 2D embedding using a Dense layer: In this case the input to the intermediate dense layer is a vector of length equal to the $num\ of\ movies$   
\end{itemize}
To determine the best approach of the two, the model is run on the IMDb feature embeddings. The results are noted in Table 1. It is seen that the first approach of flattening the 3D vector into a 2D vector gives better results for Recall@k but the second approach gives slightly better results for NDCG@k. The better results with Approach 1 are because of the information loss that occurs in Approach 2 while converting embeddings of size 3 into embeddings of size 1. Approach 1 is used for all the tasks further reported in the work because it had a better Recall@k and only a slightly worse NDCG@k.
\begin{center}
\begin{tabular}{|c|c|c|c|}\hline
Measure & Approach 1 & Approach 2 \\\hline
Eval 1 : NDCG@100 & 0.270 & \textbf{0.271}  \\\hline
Eval 1 : Recall@20 & \textbf{0.541}& 0.539\\\hline
Eval 1 : Recall@50 & \textbf{0.573} & 0.568\\\hline \hline
Eval 2 : NDCG@100 &0.181 &\textbf{0.183} \\\hline
Eval 2 : Recall@20 &\textbf{0.214} &0.211 \\\hline
Eval 2 : Recall@50 &\textbf{0.377}& 0.369\\\hline
\end{tabular}
\captionof{table}{Comparison between the approaches for handling the embedding layer}
\end{center}
\section{Visualizing Embeddings}
To better appreciate the working of a Variational Autoencoder, user and movie embeddings learned from the VAE networks are visualized. For visualizing the user embeddings, the 200-dimensional latent representation of the users obtained from Standard VAE is clustered, using k-means clustering into 10 clusters. After obtaining the cluster assignments, t-Distributed Stochastic Neighbour Embedding (t-SNE) is applied to reduce the dimensionality from 200 to 2 for the purpose of visualization. As shown in Fig. 3, users do exhibit certain patterns in their movie choices which the VAE network aims to capture. 
\begin{center}
\includegraphics[scale=0.45]{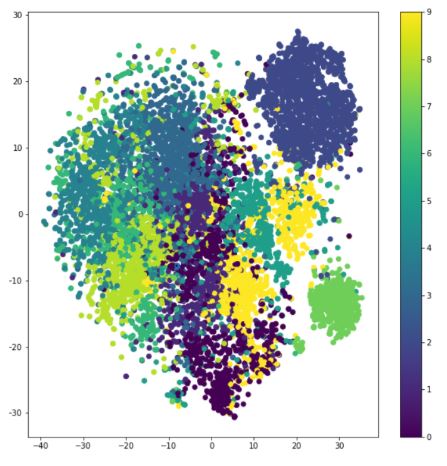}
\captionof{figure}{User Embedddings into 10 clusters}
\end{center}
Movie embeddings are also visualized in the same way as user embeddings, Fig.4. These are obtained using the M-VAE using only genres as features and are clustered into 18 clusters corresponding to the 18 genres. The visualization, encouragingly, shows a visible clustering. 
\begin{center}
\includegraphics[scale=0.3]{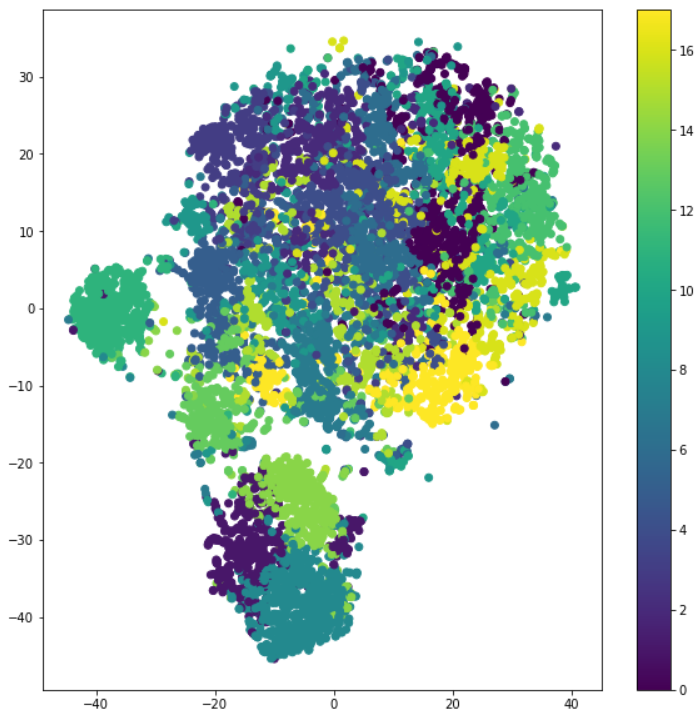}
\captionof{figure}{Movie Embedddings into 18 clusters learned using genres}
\end{center}

\section{Results and Analysis}

\begin{table*}
\begin{tabular}{|c|c|c|c|c|c|}\hline
Measure & Standard-VAE & H-VAE (Random) & H-VAE (Genre)& H-VAE (Genome)&H-VAE (IMDb) \\\hline
Eval 1 : NDCG@100 & 0.267 &0.171 & 0.249 & 0.270 & \textbf{0.271}  \\\hline
Eval 1 : Recall@20 & 0.534 &0.297 & 0.501 & 0.537 &  \textbf{0.541}  \\\hline
Eval 1 : Recall@50 & 0.562 &0.297 & 0.531 & 0.566 &  \textbf{0.572}  \\\hline \hline
Eval 2 : NDCG@100  & 0.155 &0.116 & 0.159 & 0.180 &  \textbf{0.181}  \\\hline
Eval 2 : Recall@20  & 0.208 &0.127 & 0.213 & 0.208 &  \textbf{0.215} \\\hline
Eval 2 : Recall@50  & 0.368 &0.212 & 0.368 & 0.369 &  \textbf{0.377} \\\hline
\end{tabular}
\captionof{table}{Performance of Hybrid-VAE using different feature sets compared with Standard-VAE}
\end{table*}

As discussed in Section 4, the model is run on the three feature sets and the most effective feature set is determined. The results in Table 2 show that the H-VAE outperforms the Standard-VAE, thereby verifying the significance of our feature sets. The features extracted from IMDb summaries give the highest scores, followed by genome tags. Both these feature sets outperform the movie genre feature set. This shows that the genres alone, are not a powerful contextual feature to characterize movies in recommendation systems. There are certain nuances about movies that fare beyond genres and this is shown by the fact that the H-VAE with genre features as embeddings performs worse compared to the baseline Standard-VAE.

It may be possible that the features extracted from movies have no effect at all, \& the increase in scores is only because of an extra layer. To verify this, the model was trained with random embeddings. The results show that a random embeddings layer does not add information to the model and it performs poorly even in comparison to the Standard-VAE. This verifies the usefulness and the relevance of the movie embeddings considered in this work.  

It is also possible that due to the updates during training, the embeddings change significantly from initialization, making the movie feature extraction irrelevant. The visualized IMDb feature embeddings shown in Fig. 5, depict that the training procedure does not alter the embedding space by much. And hence it makes sense to keep this information which is in the form of movie embeddings.

\begin{figure}[H]
\includegraphics[scale=0.25]{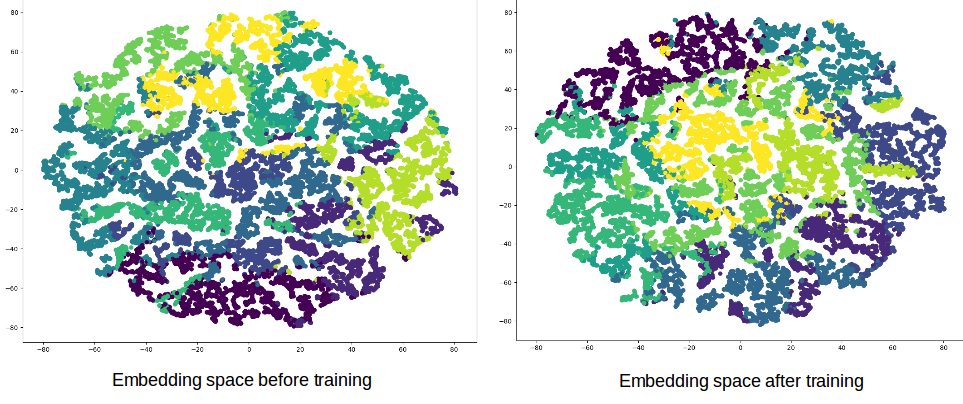}
\captionof{figure}{Comparison of embedding spaces before and after training}
\end{figure}

The IMDb features extracted are complex and contain information on the emotion and the sentiment a movie depicts. These embeddings capture the user preferences in a better manner when paired with the user rating data. The genome tags, though detailed, do not capture this sentiment/emotion and come a close second to the IMDb features. The genre feature is a high-level representation of a movie and fails to capture user preferences which appear to be more fine-grained. The genre feature set seems to add noise to the model than help in prediction. Thus, it can be stated that the IMDb feature set improves the model accuracy because it succeeds in capturing user preferences effectively, and aids in achieving a better representation of the user-item interaction.

\section{Conclusion and Future Work}
The results signify that adding context information to the item set can help increase the performance of collaborative filtering but that comes at the cost of added model complexity. Nonetheless, the proposed method gives an intuitive and flexible approach for adding high dimensional context information into a VAE network. The results also go on to show the importance and relevance of such auxiliary context information when it comes to automated recommendations. As can be noted from the results, some improvements are less than 0.01 but considering that the metrics are averaged over 10,000 users, a small increment is also indicative. For future work, the statistical significance t-test can be performed to check if the performance gain is significant or a result of noise/randomness. It will also be noteworthy to perform qualitative analysis on the learned movie embeddings to see if similar movies have similar embeddings by using an appropriate distance metric. Hyper-parameter tuning is also something that holds promise in helping to establish the robustness of the results. We publish the codebase as a public Github repository \footnote{https://github.com/kilolgupta/Variational-Autoencoders-Collaborative-Filtering\label{footnote 3}}.

\bibliographystyle{ACM-Reference-Format}
\bibliography{sample-bibliography}

\end{document}